\documentstyle[12pt]{article}
\begin{document}
\setcounter{page}{0}
\thispagestyle{empty}

\begin{center}

\vfill
{\Large
Manifestation of metric renormalization in quantum $R^2$--gravity}
\vspace{1cm}

{\large M.~Yu.~Kalmykov}
\footnote {E-mail: $kalmykov@thsun1.jinr.dubna.su$}
\vspace{1cm}

{\em Bogoliubov Laboratory of Theoretical Physics,
     Joint Institute for Nuclear Research,
     $141~980$ Dubna $($Moscow Region$)$, Russian Federation}

\vspace{1cm}

\begin{abstract}
The renormalization group method in $R^2$-gravity without matter fields
is discussed. A criterion for the existence of the renormalization
constant for the metric has been found, two-loop higher order poles have
been calculated, a relation which allows us to find the one-loop
renormalization constant of the Newtonian constant has been suggested.

\end{abstract}
\end{center}

\vfill

\pagebreak

In order to take quantum corrections into account correctly,
we should deal with the gravity which is both renormalizable and
unitary. The Einstein theory of gravity is non-renormalizable
\cite{Einstein}, whereas the theory with terms quadratic in the
curvature tensor is renormalizable in all orders of the perturbation
theory \cite{ren} but not unitary (ghosts and tachyons are present in
its spectrum). So, this $R^2$-gravity could not be accepted as a
basic theory. Nevertheless, this theory can be considered as a model
for studying quantum gravitational effects. In particular, the
renormalization group method can be used in the framework
of this theory.  However, in the framework of the standard approach
to the renormalization group method in $R^2$-gravity
\cite{RG,Avramidi1} the anomalous dimension $\gamma_G$ of the Newtonian
constant $G$ is found to be gauge dependent \cite{odin} -- which contradicts to
the quantum field theory. The introduction of the effective Vilkovisky-
DeWitt action also does not solve this problem because in this case
$\gamma_G$ becomes dependent on the configurational space metric.
A new way based on the introduction of the non-zero renormalization constant
for the metric has been recently proposed \cite{MKL} to solve the problem of
the gauge dependence of $\gamma_G$. Not only $\gamma_G$ is gauge independent
in the approach suggested, but also the renormalization constant
for the Newtonian constant G equals zero in the space-time without
a boundary in all orders of the perturbation theory both without and with
massless fields interactions. In the current paper we discuss in detail the
renormalization group formalism for $R^2$-gravity without matter fields
where the renormalization constants for the metric and the Newtonian
constant are taken into account. Two-loop higher order poles have been
calculated, a criterion for the existence of the renormalization
constant for the metric has been proposed, a relation for the definition
of the one-loop renormalization constant of the Newtonian constant G has
been found.

We use the following  notations:

$$
R^\sigma _{~\lambda  \mu \nu } = \partial_\mu \Gamma^\sigma
_{~\lambda \nu } - \partial_\nu \Gamma^\sigma _{~\lambda \mu } +
\Gamma^\sigma_{~\alpha \mu } \Gamma^\alpha_{~\lambda  \nu } -
\Gamma^\sigma_{~\alpha  \nu } \Gamma^\alpha_{~\lambda \mu },~~~~~
R_{\mu \nu } = R^\sigma_{~\mu \sigma \nu },
$$

$$
R =  R_{\mu \nu } g^{\mu \nu },
~~~~~c = \hbar = 1,
~~~~~ (g) = det(g_{\mu \nu }),
~~~~~\varepsilon = \frac{4-d}{2}, $$

\noindent
where $\Gamma^\sigma_{~\mu \nu } $ is the Riemann
connection, and $W^2 = R_{\mu \nu}^2 - \frac{1}{3}R^2$.
The space-time we work in is topologically trivial,
without a boundary, the Euler number equals zero. Thus,
we have the right to use the relation:
$R_{\mu \nu \sigma \lambda}^2 = 4 R_{\mu \nu}^2 - R^2$.

First of all, we should obtain the complete set of the renormalization
group equations in $R^2$-gravity. Two types of such equations are usually
needed:

\begin{enumerate}
\item
relations between counterterms and renormalization group functions;
\item
relations between poles of different order.
\end{enumerate}

\noindent
Nonpolynomiality of the counterterms with respect to the Lagrangian's
parameters does not permit us to use the standard renormalization group
equations. Fortunately, the necessary equations can be
easily obtained from basic principles of the renormalization group
method in a renormalizable theory \cite{min}.
Let $\{\lambda_i \}$ be a set of dimensionless (in 4-dimensional space-time)
coupling constants in a renormalizable theory. The dimension of the
bare coupling $\lambda_i^B$ in ${\bf d}$-dimensional space-time is
equal to $\rho_i \varepsilon$ ($\rho_i$ is the number).  The
renormalized couplings $\{ \lambda_j \}$ are known to be dimensionless.
In the MS-scheme in the dimensional regularization the bare
coupling constants are connected with the renormalized ones by the relation:

\begin{equation}
\lambda_j^B = (\mu^2)^{\rho_j \varepsilon} \left(
\lambda_j + \sum_{n=1} \frac{\delta Z^{(n)}_j
(\lambda_i)}{\varepsilon^n} \right),
\label{rg1}
\end{equation}

\noindent
where $\mu^2$ is the renormalization mass parameter, and
$\delta Z^{(n)}_i$ are the corresponding counterterms.
Any change of the parameter $\mu^2$ is compensated by
the appropriate change of the couplings $\{\lambda_i \}$, so that
the bare couplings $\{\lambda_i^B \}$ leave to be unchanged:
$$
\mu^2 \frac{d}{d \mu^2} \lambda_j^B = 0.
$$

\noindent
After differentiating Eq.(\ref{rg1}) with respect to $\mu^2$, and
introducing the definition

$$
\left. \mu^2 \frac{\partial}{\partial \mu^2} \lambda_j
\right|_{\lambda_j^B} = - \rho_j \varepsilon \lambda_j +
\beta_j(\lambda_i),
$$

\noindent
we get

\begin{equation}
\beta_i = \left( \sum_j \rho_j \lambda_j \frac{\partial }{ \partial
\lambda_j} - \rho_i \right) \delta Z^{(1)}_i,
\label{rg3}
\end{equation}

\noindent
and

\begin{equation}
\left( \sum_j \rho_j \lambda_j \frac{\partial }{ \partial
\lambda_j} - \rho_i \right) \delta Z^{(n)}_i = \sum_j \beta_j
\frac{\partial}{\partial \lambda_j} \delta Z^{(n-1)}_i.
\label{rg4}
\end{equation}

Let us construct the renormalization group relations for the dimensional
parameters (masses, fields). In this case the bare parameters
$\{ \kappa^B_j \}$ are related to the renormalized parameters
$\{ \kappa_j \}$ as:

$$
\kappa_j^B = \kappa_j
\left( 1 + \sum_{n=1} \frac{\delta Z_j^{(n)}}{\varepsilon^n} \right).
$$

\noindent
The invariance of bare parameters with respect to the scale renormalization
$\mu^2$ is required:
$\left( \mu^2 \frac{d}{d \mu^2} \kappa_j^B = 0 \right), $
and provided with
$$\left. \mu^2 \frac{\partial}{\partial \mu^2} \kappa_j
\right|_{\kappa_j^B} =  \gamma_j \kappa_j,$$

\noindent
we get

\begin{equation}
\gamma_i = \sum_j \rho_j \lambda_j \frac{\partial }{ \partial
\lambda_j} \delta Z_i^{(1)},
\label{rg6}
\end{equation}

\noindent
and

\begin{equation}
\sum_j \rho_j \lambda_j \frac{\partial }{ \partial
\lambda_j} \delta Z^{(n+1)}_i =
\left( \sum_j \beta_j
\frac{\partial}{\partial \lambda_j}  + \gamma_i \right)
\delta Z_i^{(n)}.
\label{rg7}
\end{equation}

\noindent
The  Eqs. (\ref{rg3}) - (\ref{rg7}) are valid for any renormalizable
theory.  In particular, the renormalization group functions are
defined there only by first order poles of the
corresponding renormalization constants. The other feature
is that the recurrent relations (\ref{rg4}), (\ref{rg7}) give
the poles of high order so that only the independent function
has the first order pole's coefficients.

Let us consider now $R^2$-gravity with the Lagrangian

\begin{equation}
L_{GR} = \Big(
\frac{1}{\lambda} W^2 - \frac{\omega}{3\lambda} R^2
- \frac{R}{k^2} + \frac{2 \sigma}{k^4} \Big) \sqrt{-g},
\label{theory}
\end{equation}

\noindent
where $k^2 = 16 \pi G$, $\lambda, \omega$, and $\sigma$ are the
constants. The dimensions of couplings in $R^2$-gravity are the
following: $\rho_\lambda = \rho_\sigma = 1, \rho_\omega  = 0$.
The $\beta$-functions in $R^2$-gravity are then defined as:

\begin{eqnarray}
\beta_{\{ \lambda, \sigma \}} & = &
\left( \lambda \frac{\partial}{\partial \lambda}
+ \sigma \frac{\partial}{\partial \sigma} - 1 \right)
\delta Z^{(1)}_{\{ \lambda, \sigma\} },
\nonumber \\
\beta_\omega & = &
\left( \lambda \frac{\partial}{\partial \lambda}
+ \sigma \frac{\partial}{\partial \sigma} \right)
\delta Z^{(1)}_\omega,
\label{rg5}
\end{eqnarray}

\noindent
where the brackets $\{ \lambda,\sigma \}$ mean that the corresponding
equations should be applied to $\lambda$ and $\sigma$ separately. In
order to obtain the anomalous dimension of the gravitational constant
$G$ and the metric field $g_{\mu \nu}$, we use Eq.(\ref{rg6}):

\begin{eqnarray}
\gamma_{\{ G, g\}} & = & \left(
\lambda \frac{\partial}{\partial \lambda} + \sigma
\frac{\partial}{\partial \sigma} \right)
\delta Z^{(1)}_{\{ G, g\} }.
\label{rg8}
\end{eqnarray}

\noindent
These equations (\ref{rg5}), (\ref{rg8})
have a nice feature: there is no derivative over $\omega$
(this parameter enters the one-loop counterterms as a nonpolynom) in
definition of both the $\beta$-functions and the anomalous dimensions.
Eqs.(\ref{rg4}),(\ref{rg7}) have the following form:

\begin{eqnarray}
&&
\left( \lambda \frac{\partial}{\partial \lambda}
+ \sigma \frac{\partial}{\partial \sigma} -1 \right)
\delta Z^{(n)}_{\{\lambda, \sigma \}} =
\left( \beta_\lambda \frac{\partial}{\partial \lambda}
+ \beta_\omega \frac{\partial}{\partial \omega}
+ \beta_\sigma \frac{\partial}{\partial \sigma}
\right) \delta Z^{(n-1)}_{\{\lambda, \sigma \}},
\nonumber \\
&&
\left( \lambda \frac{\partial}{\partial \lambda}
+ \sigma \frac{\partial}{\partial \sigma} \right)
\delta Z^{(n)}_\omega
=
\left( \beta_\lambda \frac{\partial}{\partial \lambda}
+ \beta_\omega \frac{\partial}{\partial \omega}
\right) \delta Z^{(n-1)}_\omega,
\nonumber \\
&&
\left( \lambda \frac{\partial}{\partial \lambda}
+ \sigma \frac{\partial}{\partial \sigma} \right)
\delta Z^{(n)}_{\{G, g\}} =
\left( \beta_\lambda \frac{\partial}{\partial \lambda}
+ \beta_\omega \frac{\partial}{\partial \omega}
+ \beta_\sigma \frac{\partial}{\partial \sigma}
+ \gamma_{\{G, g\}} \right) \delta Z^{(n-1)}_{\{G, g\}}.
\label{recurrence}
\end{eqnarray}

Let us calculate the one-loop renormalization group functions in
$R^2$-gravity by means of Eqs.(\ref{rg5},\ref{rg8}).
For this purpose the first things what we should find are the
corresponding renormalization constants. Let the tensor density
$g_{\mu \nu}^\ast = g_{\mu \nu} (-g)^r$ (where $r \neq - 1/4$) play
the role of a dynamical variable \footnote{The example of how to
calculate the one-loop counterterms in arbitrary parametrization in
gravity is given in Ref.~\cite{MKL2}.}. The one-loop counterterms in
the background field method have the form:

$$
\Gamma_{div}^{GR} = \frac{1}{\varepsilon}
\int
\left(\theta_2 W^2 + \frac{\theta_3}{3} R^2
+ \theta_4 \frac{R}{k^2} + \frac{\theta_5}{k^4} \right) \sqrt{-g}
~d^4 x.
$$

In the theory (\ref{theory}) the following renormalization
constants may arise within the background field method:
for the physical parameters $Z_\lambda, Z_\omega, Z_G, Z_\sigma$,
for the background metric field $Z_g$, and for quantum and ghost fields
$Z_{h}$ and $Z_{gh}$, respectively. All the ultraviolet singularities
of the theory can be absorbed into these constants. In the MS-scheme
all $Z_i$ constants contain poles only in $\varepsilon$.
Following the arguments of \cite{Abbott} it can be shown that the
renormalization of quantum and ghost fields in the background field
method is not essential to the one-loop background Green
functions' calculation. Supposing that after the multiplicative
redefinition of the metric and parameters

\begin{eqnarray}
g_{\mu \nu}^* & \to &
\left(1 + \frac{\delta Z_{g,s}^{(1,1)}}{\varepsilon} \right) g_{\mu \nu}^{*},
~~~~~G \to
G_B = \left( 1 + \frac{\delta Z_G^{(1,1)}}{\varepsilon} \right) G,
\nonumber \\
\lambda & \to &
\left(\lambda + \frac{\delta Z_\lambda^{(1,1)}}{\varepsilon} \right),
~~~~~
\omega \to
\left(\omega + \frac{\delta Z_\omega^{(1,1)}}{\varepsilon} \right),
~~~~~
\sigma \to
\left(\sigma + \frac{\delta Z_\sigma^{(1,1)}}{\varepsilon} \right),
\nonumber
\end{eqnarray}

\noindent
the one-loop background Green functions obtained from the effective
action $\Gamma = \Gamma_B^{GR} - \Gamma_{div}^{GR}$ should be
finite at $\varepsilon \to 0$, we have the following system of
equations for $\delta Z_i$:

\begin{eqnarray}
\delta Z_\lambda^{(1,1)} & = & - \theta_2 \lambda^2,
\label{Zl} \\*
\delta Z_\omega^{(1,1)} & = &
- \left( \theta_3 + \theta_2 \omega \right) \lambda,
\label{Zo} \\*
\delta Z_\sigma^{(1,1)}  & = & \frac{\theta_5 + 4
\theta_4 \sigma }{2},
\label{Zs} \\*
\delta Z_G^{(1,1)}  & - & \frac{1}{s} \delta Z_{g,s}^{(1,1)}
=  \theta_4,
\label{Z}
\end{eqnarray}

\noindent
where $s \equiv 4r+1 \neq 0$.
At the one-loop level the coefficients $\theta_2, \theta_3$, and
the combination $\theta_5 + 4 \sigma \theta_4$ are independent on
the gauge and the parametrization off-shell \cite{Avramidi}. By using the
relations (\ref{Zl}) - (\ref{Zs}), we get that the one-loop
renormalization constants and the corresponding $\beta$-functions
are also gauge and parametrization independent.

The one-loop counterterms in Ref.\cite{Avramidi1}
have been used to calculate the $\beta$-functions:

\begin{eqnarray}
\beta_\lambda^{(1)} & = &
- \frac{1}{16 \pi^2} \frac{133}{20} \lambda^2
= \delta Z_\lambda^{(1,1)},
\nonumber \\
\beta_\omega^{(1)} & = &
- \frac{1}{16 \pi^2} \left(
\frac{5}{3} \omega^2 + \frac{183}{20} \omega + \frac{5}{24}
\right) \lambda
= \delta Z_\omega^{(1,1)},
\nonumber \\
\beta_\sigma^{(1)} & = & \frac{\lambda}{32 \pi^2} \left(
\frac{5}{4} \lambda + \frac{\lambda}{16 \omega^2}
+ \frac{20}{3} \omega \sigma + 5 \sigma
- \frac{1}{6} \frac{\sigma}{\omega} \right)
= \delta Z_\sigma^{(1,1)}.
\label{beta}
\end{eqnarray}

\noindent
The one-loop $\beta$-functions (\ref{beta}) obtained by using (\ref{rg5})
turn out to be equal to the standard values \cite{RG,Avramidi1}. The one-loop
anomalous dimensions satisfy the following equations (in the trivial
parametrization $s = 1$ ):

\begin{equation}
\gamma_G^{(1)}- \gamma_{g,1}^{(1)} =  \frac{1}{16 \pi^2}
\left( \frac{5}{3} \omega - \frac{13}{12} - \frac{1}{8 \omega}
\right) \lambda = \delta Z_G^{(1,1)} - \delta Z^{(1,1)}_{g,1}.
\label{anom}
\end{equation}

\noindent
We should pay attention to the fact that the $\theta_4$ is gauge and
parametrization dependent coefficient \cite{Avramidi}. G is a quantity
which can be measured in experiments and enters the classical
gravitational potential's definition \cite{stelle}. The loop
corrections to the gravitational potential are proportional to
the $\gamma_G$-function. This results in the gauge and
parametrization dependence of physical quantities. Therefore,
the $\gamma_G$ must be gauge and parametrization independent quantity.
This can be reached only by introduction of the non-zero gauge and
parametrization dependent function $\gamma_g$
(see Eq. (\ref{anom})), so that only unphysical parameter
-- the anomalous dimension of field -- depends on the gauge and
parametrization. Fortunately, this does not contradict
to the basics of the quantum field theory. From Eq. (\ref{Z})
it is easy to have:

\begin{equation}
\frac{ \delta Z_{g,s_1}^{(1,1)} }{s_1}
- \frac{ \delta Z_{g,s_2}^{(1,1)} }{s_2}
=  \theta_4^{(s_2)} - \theta_4^{(s_1)},
\label{one-loop}
\end{equation}

\noindent
where $\theta_4^{(s)}$ is the counterterm obtained in the parametrization
$s$. If the renormalization constant for the metric equals zero,
then $\theta_4^{(s)}$ does not depend on $s$, and (\ref{one-loop})
is identically equal to zero. So, the dependence of $\theta_4$ on
the parametrization (not on the gauge!) may be a criterion of the
existence of the renormalization constant for the metric field.

By the same way the renormalization group equations can be found
in the conformal parametrization where the fields $\psi_{\mu \nu} =
g_{\mu \nu} (-g)^{-\frac{1}{4}}$, and $\pi = (-g)^{\frac{m}{4}}$,
$m \neq 0,~~ det \psi_{\mu \nu} = 1$ act as independent dynamical
variables. In this case there are two different renormalization
constants $Z_\psi, Z_\pi$ (as well as the two anomalous dimensions
$\gamma_\psi$ and $\gamma_\pi$ appear): for the field $\psi$
(which concerns the traceless part of the metric tensor)
and $\pi$ (which is related to its trace), respectively.
Eq.(\ref{rg8}) should be replaced with:

$$
\gamma_{\{ \psi, \pi \}} =  \left(
\lambda \frac{\partial}{\partial \lambda} + \sigma
\frac{\partial}{\partial \sigma} \right)
\delta Z^{(1)}_{\{\psi, \pi\}}.
$$

\noindent
Provided with the gauge and parametrization independence of
$\delta Z_\lambda^{(1,1)}$ and $\theta_2$, we get
$\delta Z_\psi^{(1,1)} = 0$.
So, we see that at the one-loop level only the conformal mode of the metric
must be renormalized.
The old solutions (\ref{Zl})--(\ref{Zs}) are obtained once more, in
the solution (\ref{Z}) we should replace $s \to m, g \to \pi$.

The recurrent relations and the low order values of the renormalization
functions allow calculations to be made of their higher order contribution
and to calculate the coefficients at high order poles of the two-loop
counterterms.
Let the two-loop counterterms in the background-field method have
the form:

\begin{eqnarray}
\Gamma_{div}^{(2)} & = & \int \Biggl[
\frac{1}{\varepsilon^2}
\left(\theta_2^{(2,2)} W^2 + \frac{\theta_3^{(2,2)}}{3} R^2
+ \theta_4^{(2,2)} \frac{R}{k^2} + \frac{\theta_5^{(2,2)}}{k^4}
\right)
\nonumber \\*
&& + \frac{1}{\varepsilon}
\left(\theta_2^{(1,2)} W^2 + \frac{\theta_3^{(1,2)}}{3} R^2
+ \theta_4^{(1,2)} \frac{R}{k^2} + \frac{\theta_5^{(1,2)}}{k^4}
\right)
\Biggr]
\sqrt{-g} ~d^4 x.
\nonumber
\end{eqnarray}

\noindent
Taking into account  Eqs.(\ref{recurrence}) with
(\ref{beta}) and (\ref{anom}), we have:

\begin{eqnarray}
\theta_5^{(2,2)} & = & \left(\frac{1}{16 \pi^2} \right)^2
\Biggl[
\lambda^2 \left(
-  \frac{1}{32 \omega^2}
+ \frac{31}{72} \frac{\sigma}{\omega}
- \frac{763}{9} \sigma \omega
- \frac{50}{9} \sigma \omega^2 - \frac{163}{24} \sigma
\right)
\nonumber \\
& + & \lambda^3 \left(
- \frac{259}{24} + \frac{7}{192} \frac{\sigma}{\omega^2}
+ \frac{29}{48 \omega^2} + \frac{5}{12} \frac{\sigma}{\omega}
+ \frac{5}{12 \omega} + \frac{316}{9} \sigma \omega
+ \frac{100}{27} \sigma \omega^2 - \frac{4787}{540} \sigma
\right)
\Biggr]
\nonumber \\*
& + & \frac{\sigma \lambda }{4 \pi^2}
\left( \frac{5}{3} \omega - \frac{13}{12} - \frac{1}{8 \omega}
\right) \delta Z_G^{(1,1)},
\nonumber \\
\theta_4^{(2,2)} & = &
- \left( \frac{1}{16 \pi^2} \right)^2
\Biggl[
\frac{\lambda^3}{3}
\left(-\frac{4787}{720} + \frac{25}{9} \omega^2 + \frac{79}{3} \omega
+ \frac{5}{16 \omega} + \frac{5}{192 \omega^2} \right)
\nonumber \\*
&&
+ \frac{\lambda^2}{2}
\left( \frac{13}{12} - \frac{5}{3} \omega
+ \frac{1}{8 \omega} \right) \Biggr]
- \frac{\lambda}{16 \pi^2}
\left( \frac{5}{3} \omega - \frac{13}{12} - \frac{1}{8 \omega}
\right) \delta Z_G^{(1,1)},
\nonumber \\
\theta_3^{(2,2)} & = & -\left(\frac{1}{16 \pi^2} \right)^2
5 \lambda \left( \frac{\omega}{3} + \frac{1}{4} \right)
\left( \frac{5}{3} \omega^2 + \frac{183}{20} \omega +
\frac{5}{24}\right),
\nonumber \\
\theta_2^{(2,2)} & = & 0.
\label{two-loop2}
\end{eqnarray}

\noindent
The coefficients $\theta_2^{(2,2)}$ and $\theta_3^{(2,2)}$
are gauge and parametrization independent due to the gauge and
parametrization independence of the corresponding one-loop
renormalization constants $\{ \delta Z_i \}$.
It should be noted that the two-loop counterterms
(\ref{two-loop2}) contain the terms proportional to $1/\omega^3$.
The relations (\ref{two-loop2}) yield double profit:

\begin{itemize}
\item
gauge and parametrization independent coefficients $\theta_2^{(2,2)}$
and $\theta_3^{(2,2)}$ predicted from these relations can be used for
testing the formalism of the renormalization group method suggested
here.

\item
coefficients $\theta_4^{(2,2)}$ and $\theta_5^{(2,2)}$ can be used for
the definition of the $\delta Z_G^{(1,1)}$.
\end{itemize}

In this paper a thorough analysis of the renormalization group method in
$R^2$-gravity has been carried out. The Eqs. (\ref{rg5}),
(\ref{rg8}) between the renormalization group functions and the
renormalization constant, as well as the recurrence relation between the
corresponding pole terms (\ref{recurrence}) have been obtained.
It is noteworthy to emphasize that although the counterterms are
not polynoms in the Lagrangian's parameters, all the qualitative
properties of the renormalization group equations in this model
are the same as in the standard quantum field theory.
These equations allowed us to suggest a criterion for the existence
of the renormalization constant for the metric (Eq. (\ref{one-loop})),
to calculate two-loop higher order poles, as well as to show how
the one-loop renormalization constant for the Newtonian constant
should be found from the two-loop results ( Eq.(\ref{two-loop2}) ).

\noindent
{\bf Note added}

\noindent
When this paper has already been finished  we became aware of the paper
\cite{Pronin2}. The results of calculation of the one-loop counterterm,
in arbitrary gauge and parametrization, obtained in \cite{Pronin2}
directly confirm the presence of nonzero renormalization constant for the
metric field (see Eq. \ref{one-loop} in the current paper).

\end{document}